# Study of quasi-particle dynamics using the optical pulse response of a superconducting resonator

J. Hu,[1] Q. He,[2] F. Yu,[2] Y. Chen,[1] M. Dai,[2] H. Guan,[1] P. Ouyang,[2] J. Han,[2] C. Liu,[1] X. Dai,[1] Z. Mai,[1] X. Liu,[1] M. Zhang,[1] L. F. Wei,[2,a)] M. R. Vissers,[3] J. Gao,[3] and Y. Wang[1,a)]

[1)]*Quantum Optoelectronics Laboratory, School of Physical Science and Technology, Southwest Jiaotong University, Chengdu, Sichuan 610031, China.*
[2)]*Information Quantum Technology Laboratory, School of Information Science and Technology, Southwest Jiaotong University, Chengdu, Sichuan 610031, China.*
[3)]*National Institute of Standards and Technology, Boulder, CO 80305, USA*

[a)]Authors to whom correspondence should be addressed: qubit@swjtu.edu.cn and lfwei@swjtu.edu.cn

(Dated: 7 August 2021)

We study the optical pulse response of a superconducting half-wavelength coplanar waveguide (CPW) resonator. We apply a short optical pulse to the center strip of the CPW resonator, where the current distribution shows antinodes or nodes for different resonance modes, and measure the frequency response. We develop a time-dependent variable inductance circuit model with which we can simulate the optical pulse response of the resonator. By fitting this model to experimental data, we extract the temporal kinetic inductance variations, which directly reflects the quasi-particle recombination with time and diffusion in space. We also retrieve the spatial size of the quasi-particle distribution and the quasi-particle diffusion constant. Our study is very useful for the design of photon-counting kinetic inductance detectors and the method developed in this work provides a useful way to study the quasi-particle dynamics in the superconductor.

Superconducting micro-resonators[1] have important applications in sensitive photon detection. For example, microwave kinetic inductance detectors (MKIDs)[2–4] are made from high quality factor (high-$Q$) superconducting thin film resonators. Because MKIDs can be easily incorporated into a large array, they have been widely used in detecting weak photon flux at sub-millimeter and millimeter wavelengths for astrophysical and imaging applications[5–9]. MKIDs have also demonstrated good capability to count single photons and resolve photon energy at infrared, visible and X-ray wavelengths[10–17]. For these photon-counting applications, it is necessary to have an in-depth understanding of the optical pulse response of superconducting micro-resonators in the time domain. This is important not only for the detector design, but also for understanding the quasi-particle (QP) dynamics [18–23] and other underlying physics in superconducting thin film devices.

It is known that a non-equilibrium distribution of quasi-particles (QPs) will be quickly created (typically within a few nanoseconds) in the superconductor after the absorption of an energetic photon[24–26]. Then the QPs diffuse in the superconductor and recombine. This QP recombination process normally lasts for a few microseconds to a few milliseconds [27–33], mainly depending on the superconducting material, film thickness and substrate of choice. The temporal and spatial evolution of QP density and distribution simultaneously changes the kinetic inductance (also resistance) of the superconducting film, resulting in a pulsed change in the resonance frequency, which can be read out in a time-domain homodyne measurement.

In this paper, we study the optical pulse response of a superconducting aluminum (Al) half-wavelength coplanar waveguide (CPW) resonator. We apply a short optical pulse to the center of the CPW, where the current distribution shows antinodes or nodes for different resonance modes, and we measure the corresponding frequency responses. We develop a time-dependent variable inductance circuit model to simulate the optical pulse response of the resonator. By fitting the measured data to our model, we have retrieved the temporal kinetic inductance variation profile, which directly reflects the QP recombination process in the time domain because the kinetic inductance change is linear with the QP density in thin superconducting films[34]. We also use a method to estimate the spatial distribution of QPs with time and the QP diffusion constant, by considering the position dependent current distribution and comparing the response of odd resonance modes to even resonance modes.

As shown in Fig. 1(a), we study a half-wavelength CPW resonator with length $l$ and coupling capacitors $C_c$ on both the input and output ends. The characteristic impedance of the CPW is given by $Z_0 = \sqrt{\widetilde{L}/\widetilde{C}} \approx 50$ ohm, where $\widetilde{L}$ and $\widetilde{C}$ are the distributed inductance and capacitance per unit length respectively. At resonance frequencies $f_n = nf_1$, the microwave signal can pass through the circuit and resonance peaks can be observed in transmission ($S_{21}$) measurements. Here $n = 1, 2, ...$ is the resonance mode index and $f_1 = 1/(2l\sqrt{\widetilde{L}\widetilde{C}})$ is the fundamental resonance frequency. On the $n$-th resonance, the current amplitude distribution shows a standing-wave pattern along the CPW and follows the position-dependent relation $I(x) = I_m \sin(n\pi x/l)$, where $x$ is the position and $I_m$ is the maximum current amplitude.

It can be derived (see the supplementary material) that around the resonance frequency the half-wavelength CPW behaves like an equivalent $RLC$ parallel resonance circuit, which is shown in Fig. 1(b). In the following we define the effective lumped-element capacitance and inductance as $L = 2l\widetilde{L}/\pi^2$ and $C = l\widetilde{C}/2$. Then the equivalent circuit parameters (in Fig. 1(b)) for the $n$-th resonance mode are given by:

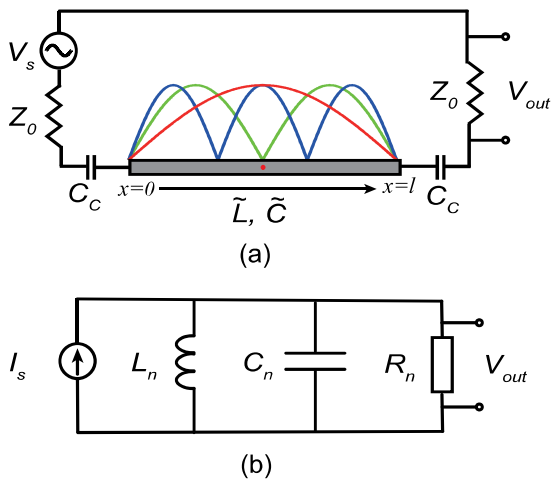

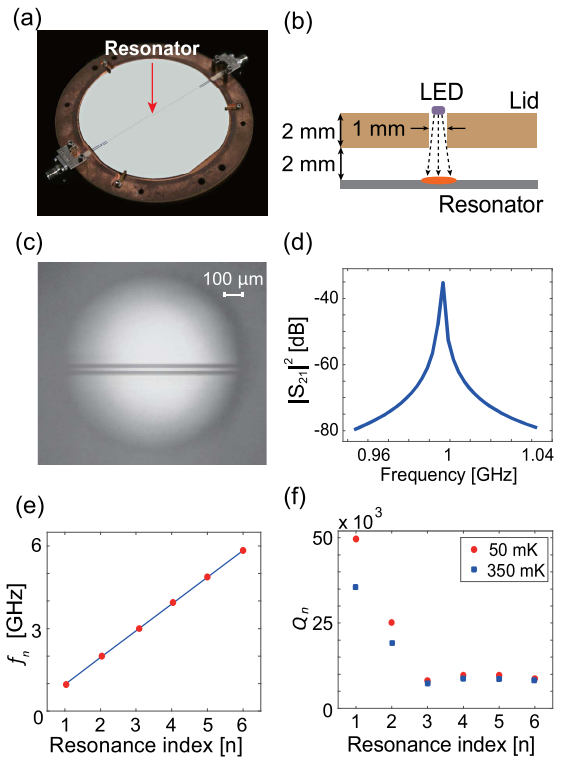

FIG. 1. (a) A half-wavelength CPW resonator with coupling capacitors to input/output port. The three dotted curves (red, green, blue) show the current amplitude distribution for the first three resonance modes ($n = 1, 2, 3$) respectively. The red point represents the center position of the CPW that is illuminated by the LED. (b) On the $n$-th resonance, the CPW is equivalent to a lumped-element $R_n L_n C_n$ resonance circuit with a current source $I_s$.

FIG. 2. (a) A photo of the Al CPW half-wavelength resonator mounted in a circular sample holder (made from oxygen-free copper) with input/output SMA connectors. (b) A small hole (about 1 mm in diameter) is drilled on an aluminum lid (about 2 mm thick) as a collimator. The collimator hole is aligned with the center position of the resonator. The vertical distance between the aluminum lid and the superconducting film is about 2 mm. A 1550 nm LED (about 0.8 mm in diameter) is placed on top of the collimator hole so that the LED illuminates a small area around the resonator center. The backside of the lid is painted black to minimize the light reflected and scattered inside the sample box. (c) Micrograph of the CPW transmission line through the collimator hole. The transmission line is placed in the middle of the field of view. (d) The measured transmission ($S_{21}$) of the fundamental resonance. (e) The measured resonance frequencies of the first 6 resonance peaks. (f) The total quality factor for the first 6 resonance modes at temperature 50 mK and 350 mK.

$$C_n = C \quad L_n = \frac{1}{n^2} L \quad R_n = Q_n Z_r \qquad (1)$$

where $Q_n = \omega_n R_n C_n$ is the total quality factor for $n$-th resonance mode and $Z_r = \sqrt{L/C} = 2Z_0/\pi$ is the resonator impedance. These parameters set the resonance frequencies at $f_n = 1/(2\pi\sqrt{L_n C_n}) = n f_1$, where $f_1 = 1/(2\pi\sqrt{LC})$ is the fundamental resonance frequency.

As shown in Fig. 2(a), we fabricate a $l = 59.26$ mm long half-wavelength CPW resonator, which is made from a superconducting Al film ($\approx 120$ nm thick and sheet resistance $\approx 0.8$ Ohm/□) deposited and patterned on a high-resistivity Si wafer (76.2 mm in diameter) by magnetron sputtering, standard photolithography and wet etch. The designed width of the center strip is 15 $\mu$m and the gap between the center strip and ground is 15 $\mu$m. The designed fundamental frequency is about 1 GHz. For such a thin-film CPW transmission line, one can estimate its kinetic inductance fraction[35] to be about 1.2%. The CPW resonator is coupled at each end through a small capacitance to the input/output ports and the designed coupling quality factor ($Q_c$) for $n = 1$ resonance is about $50 \times 10^3$.

The resonator device is mounted in a circular copper sample holder with input and output SMA ports. As shown in Fig. 2(b), a 1550 nm LED illuminates a small area around the center of the CPW through a collimator hole on the lid. The assembly, with the resonator wafer in the holder with lid and LED, is mounted on the mixing chamber (MC) stage in a dilution refrigerator (DR) and cooled down to a base temperature of 50 mK. We first measured the transmission $S_{21}$ of the resonator in the dark by using a vector network analyzer (VNA). The transmission peak of the first resonance is plotted in Fig. 2(d), showing the fundamental frequency is $f_1 = 0.96987$ GHz and $Q_1 = 50.1 \times 10^3$, which agree well with our design. The frequencies of the first 6 resonance peaks are shown in Fig. 2(e) and the linear spacing between the resonance frequencies agrees well with the theory $f_n = n f_1$. Fig. 2(f) shows the total quality factor $Q_n$ for the first 6 resonance modes at temperatures 50 mK and 350 mK respectively.

For the optical response measurement, we use a function generator at room temperature to drive the LED to emit optical pulses with a short width of 200 ns at a repetition rate of 120 Hz. The optical pulse width is short enough to generate excess quasi-particles (due to external pair breaking) while avoiding producing thermally excited quasi-particles[34,36]. As shown in Fig. 3(a), a standard homodyne scheme is used to read out the resonator responses. We probe the resonators at the first 6 resonance frequencies. The microwave power is chosen to be far below bifurcation power to avoid strong non-linear effects in the resonator, and the resonator responses show little microwave power dependence in the weak power regime. For each optical pulse, the corresponding response of the detector is digitized at a sampling rate of 2.5 Ms/s. Fig. 4(a) shows a typical resonator response in the complex IQ plane. The raw data are then converted to the frequency and dissipation responses. We only further analyze the frequency response data to retrieve the kinetic inductance effects.

The fractional frequency shift $\delta f_n/f_n$ at 50 mK for the first 6 resonance modes are plotted as blue curves in Fig. 4(b). Note that the frequency responses for resonance modes $n = 1, 3, 5$ are much larger than the responses for modes $n = 2, 4, 6$. This is expected because the illuminated region is an antinode with maximum current amplitude for odd resonance modes and a node with minimum current amplitude for even resonance modes. Although the same optical pulse is applied to the current antinodes, the $n = 1, 3, 5$ resonance responses measured at the mixer output exhibit a few different features in several aspects, including response times, decay rates and pulse heights, due to different $Q_n$ and resonator bandwidth. It is known that for small perturbations in impedance, the resonator circuit acts as a low-pass filter with a bandwidth equal to the resonator's bandwidth $f_n/2Q$[37]. In the adiabatic limit ( slow change), the frequency response will be proportional to the impedance change. In order to study the resonator response in general situations (without approximation on amplitude and rate of impedance change), we develop a variable inductance circuit model, which can simulate the resonator response for any impedance variations. Therefore our model can serve as a more general and quantitative approach to analyze the optical pulse response of microwave resonator.

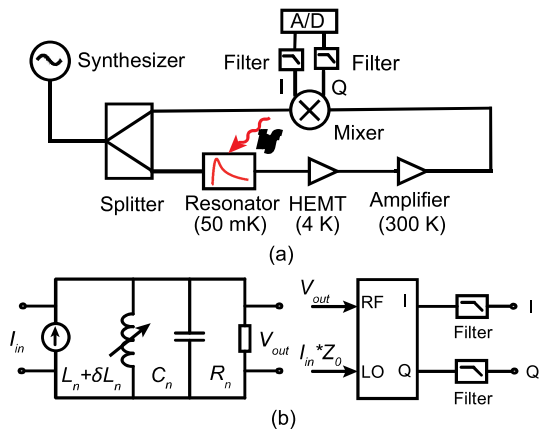

FIG. 3. (a) Pulse response measurement circuits. (b) Variable inductance circuit method to simulate the optical pulse response of the resonator.

Our variable inductance circuit model is shown in Fig. 3(b), which is based on the equivalent resonance circuit in Fig. 1(b). In simulation, the voltage across the variable inductor satisfies the Faraday's law, i.e., $V(t) = L(t) \mathrm{d}(I(t))/\mathrm{d}t + I(t)\mathrm{d}(L(t))/\mathrm{d}t$. Refer to the supplementary material for more details on this model.

Due to the photon induced changes in the QP density, the perturbed distributed inductance $\delta\widetilde{L}(x,t)$ of the superconducting CPW will vary as a function of position and time. From a modal analysis[38] one can derive that the equivalent inductance variations are weighted by the square of the current distribution along the CPW and the time-dependent fractional inductance change for $n$-th resonance mode is given by:

$$\frac{\delta L_n(t)}{L_n} = \frac{2}{l\widetilde{L}} \int_0^l \delta\widetilde{L}(x,t) \sin^2 \frac{n\pi x}{l} \mathrm{d}x \qquad (2)$$

However, $\delta\widetilde{L}(x,t)$ is usually hard to know because of the combined effects of QP recombination and diffusion[17,18]. In order to disentangle these two effects, we make a simple but reasonable assumption that $\delta\widetilde{L}(x,t)$ can be approximately determined by pure recombination weighted by a Gaussian distribution factor, i.e., $\delta\widetilde{L}(x,t) = \delta\widetilde{L}(t)\frac{\sigma_0}{\sigma(t)}e^{-\frac{(x-x_0)^2}{2\sigma^2(t)}}$ (see supplementary material for more details). Here $\delta\widetilde{L}(t)$ describes the pure QP recombination effect without QP diffusion. Note that $\delta\widetilde{L}(t)$ is usually not an exponential decay. We also assume a Gaussian distribution of QPs and the distribution width $\sigma(t)$ increases with time due to QP diffusion along the CPW. $\sigma_0 = \sigma(t = 0)$ is the initial size of the light spot. Because of the size of the collimator hole, $\sigma_0$ should be about 1 mm, which is much shorter than the microwave wavelength (a few centimeters) but longer than the QP diffusion length in Al (a few hundred micrometers supposing the QP lifetime is a few tens of microseconds). Then Eqn. (2) can be rewritten as:

$$\frac{\delta L_n(t)}{L_n} = \frac{\delta\widetilde{L}(t)}{\widetilde{L}} \frac{2}{l} \int_0^l \frac{\sigma_0}{\sigma(t)} e^{-\frac{(x-x_0)^2}{2\sigma^2(t)}} \sin^2 \frac{n\pi x}{l} \mathrm{d}x \qquad (3)$$

which can be further expressed in a compact form:

$$\frac{\delta L_n(t)}{L_n} = \frac{\delta\widetilde{L}(t)}{\widetilde{L}} A_n(t) \qquad (4)$$

From Eqn. (4), one can see that the variation in the equivalent inductance for the $n$-th resonance mode is equal to the relative change in the distributed inductance $\delta\widetilde{L}(t)/\widetilde{L}$ multiplied by a time-dependent factor $A_n(t)$. As mentioned above, $\delta\widetilde{L}(t)/\widetilde{L}$ describes the temporal variations in the distributed inductance, which can be considered as a direct reflection of the QP creation and recombination process. In theory, $\delta\widetilde{L}(t)/\widetilde{L}$ should be the same for all the resonance modes because the same optical pulse is applied to the same position, and the photon-device coupling is fixed. $A_n(t)$ is a resonance mode dependent coefficient, which takes into account the spatial distribution of QPs and current distribution. For a resonant microwave input signal and a given variable inductance $\delta L_n(t)/L_n$, one can calculate the dynamical output voltage, which is then mixed with the input signal (reference signal) and filtered by a low pass filter. In this way, the optical pulse response of the resonator can be simulated.

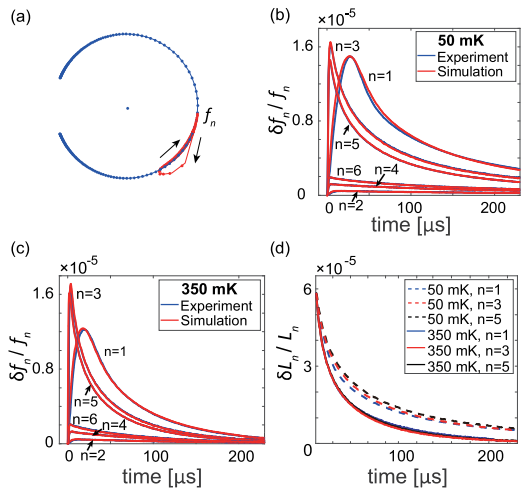

FIG. 4. (a) Optical pulse response (red curve) and resonance circle (blue curve) in the I/Q complex plane. (b) Experimental frequency responses (blue curves) and simulated frequency responses (red curves) for the first 6 resonances at 50 mK. (c) Experimental frequency responses (blue curves) and simulated frequency responses (red curves) for the first 6 resonances at 350 mK. (d) The extracted $\delta L_n/L_n$ ($n = 1, 3, 5$) at 50 mK and 350 mK, showing good agreements.

On the other hand, we can also use an iteration method (refer to supplementary material for more details) to reconstruct the inductance pulse $\delta L_n(t)/L_n$ from experimental frequency response data. Specifically, we use the falling stage of experimental frequency response $\delta f_n/f_n$ ($n = 5$) as the initial solution for $\delta L_n(t)/L_n$, because the $n = 5$ resonance has higher resonance frequency and lower $Q$. Therefore the $n = 5$ resonance has a larger bandwidth, a shorter response time and its frequency response $\delta f_n(t)/f_n$ is more similar to $\delta L_n(t)/L_n$ in shape. For the rising stage of $\delta L_n/L_n$, we assume a linear slope to a maximum for the first 200 ns. With this trial solution, we can simulate the corresponding pulse response for all the resonance modes in the time domain. The deviation between the simulated results and experimental data is used to correct the input $\delta L_n(t)/L_n$. After a few iterations, one can find a best match between the simulated response and experimental response. Correspondingly, the best solution $\delta L_n(t)/L_n$ for all resonances can be obtained.

Fig. 4(b) plots the experimental responses and simulated responses for all the resonances at 50 mK, showing that the resonator responses can be well simulated with experimental resonance parameters. The corresponding $\delta L_n/L_n$ (magnitude is normalized to $\delta L_1/L_1$) for $n = 1, 3, 5$ are plotted as dashed curves in Fig. 4(d). One can see that the shapes of extracted $\delta L_n/L_n$ are similar. For odd resonance modes $n = 1, 3, 5$, the illuminated position is an antinode with maximum current amplitude and a flat current distribution, so that $A_n(t)$ is not sensitive to QP diffusion and should be approximately the same for $n = 1, 3, 5$ resonances according to Eqn. (4). Therefore the profile of $\delta L_n/L_n$ mainly depends on the pure recombination $\delta \widetilde{L}(t)/\widetilde{L}$, which is not resonance mode dependent. In other words, with our method one can extract the real quasi-particle dynamics by removing the resonance circuits effects. For comparison, we also plot the experimental responses and simulated responses for all the resonances at 350 mK in Fig. 4(c). At higher temperatures, the $Q$ of all the resonances are lowered (see Fig. 2(f)). Therefore the pulse responses at 350 mK show some different features (e.g., shorter response time, faster relaxation and different pulse height), compared to the results at 50 mK. Fig. 4(c) shows that the experimental results can be also well simulated at 350 mK. The extracted and normalized $\delta L_n/L_n$ for $n = 1, 3, 5$ at 350 mK are plotted as solid curves in Fig. 4(d), which are also close to each other.

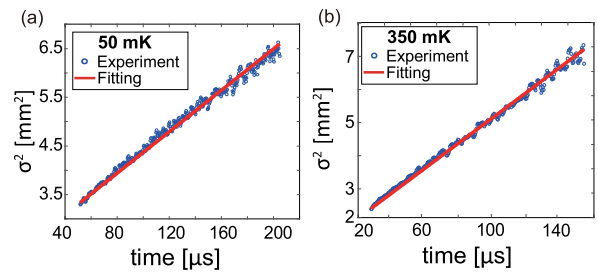

FIG. 5. (a) The extracted $\sigma^2(t)$ and a linear fitting with time at 50 mK. (b)The extracted $\sigma^2(t)$ and a linear fitting with time at 350 mK. The data at longer time shows larger fluctuations due to decreasing signal to noise ratio.

For even resonance modes ($n = 2, 4, 6$), the optical pulse responses can also be well simulated, but requiring a much slower falling stage of $\delta L_n/L_n$ than odd resonance modes. This can be mainly attributed to QP diffusion around the current nodes where current amplitude increases rapidly away from the nodes. By taking advantage of this effect, one can estimate the QP diffusion constant from experiments. Based on Eqn. (3) and Eqn. (4), the ratios of $\delta L_n/L_n$ at time $t$ between odd resonance modes and even resonance modes depend sensitively on the QP width $\sigma(t)$. Note that the ratios of $\delta L_n/L_n$ between even resonance modes or between odd resonance modes are not sensitive to $\sigma(t)$ because of the similar position dependent current distribution. By calculating the ratios of $\delta L_n/L_n$ ($n = 1, 3, 5$) to $\delta L_n/L_n$ ($n = 2, 4, 6$) and averaging the results, we can get $\sigma(t)$ as a function of time. For $t = 0$, the initial width $\sigma_0$ of QP distribution (without diffusion) is determined to be 1.27 mm for 50 mK and 1.33 mm for 350 mK, which are both consistent with the size of the collimator hole. For a one-dimensional diffusion with Gaussian shape and satisfying the Fick's law, we have $\sigma^2(t) = \sigma_0^2 + 2Dt$, where $D$ is the QP diffusion constant. Fig. 5(a) and Fig. 5(b) plot $\sigma^2(t)$ with time for 50 mK and 350 mK respectively, showing good linearity. From the slopes, we can estimate that $D = 103$ cm$^2$/s at 50 mK and $D = 187$ cm$^2$/s at 350 mK in our superconducting aluminum films. These values are relatively larger than the QP diffusion constants reported in the literature[27–29], ranging from 22.5 cm$^2$/s to 100 cm$^2$/s.

Here we discuss the uncertainty in estimating the diffusion constant. Actually, we can calculate the ratio of $\delta L_n(t)/L_n(t)$ between only one pair of odd and even resonance modes to estimate the QP distribution width $\sigma(t)$ at time $t$. There are 3 odd resonance modes ($n = 1,3,5$) and 3 even resonance modes ($n = 2,4,6$), therefore there are 9 combinations of odd and even resonances from which we can independently estimate the diffusion constant $D$ and the initial QP distribution width $\sigma_0$. The detailed results are summarized in TABLE S1 in the supplementary material. At 50 mk, $D$ is estimated to be from 48 cm$^2$/s to 164 cm$^2$/s and $\sigma_0$ is estimated to be from 0.85 mm to 1.69 mm. Similarly, at 350 mK, $D$ is estimated to be from 85 cm$^2$/s to 250 cm$^2$/ and $\sigma_0$ is estimated to be from 0.97 mm to 1.76 mm. We think there are at least two reasons which may lead to a large uncertainty in estimating $D$ and a higher $D$ value than the typical QP diffusion constants in Aluminum film reported in the literature. First, we did not consider the effects that recombination time depends on QP density in Eqn. (4), which simply assumes the same $\widetilde{\delta L}(t)$ for all resonance modes. The ratio between odd and even resonance modes (e.g., $\frac{\delta L_1(t)}{L_1}/\frac{\delta L_2(t)}{L_2}$) generally decreases with time, and this decrease is all attributed to the diffusion effects (weighted by current distribution). In other words, the contribution of the fast recombination at the illumination center for resonance 1 is underestimated. Therefore we see a wider QP distribution width $\sigma(t)$ (faster diffusion) than its true value at time $t$ and $D$ is overestimated. The estimated $D$ will be slightly smaller if we use data at longer time, which is an evidence that the recombination time is QP density dependent. Second, due to the high reflectivity of Al film and larger collimator hole diameter ($\approx 1$ mm), the incident photons will be reflected and scattered inside the sample box. This may cause the QPs to have a non-Gaussian distribution, which will complicate the resonator response. These will be studied further in our future experiments.

In conclusion, we have studied the optical pulse response of a superconducting half-wavelength CPW resonator. We demonstrate a variable inductance circuit method which can well simulate the frequency response of the resonator in the time domain. With this method, we can remove the resonance circuits effects on the resonator response and determine the temporal kinetic inductance variations (i.e., the real quasi-particle recombination process). We also report a method to estimate the time-dependent spatial distribution of quasi-particles. This method uses the sensitivity of the resonator response to the position-dependent current distribution, and one can compare the resonator response between odd and even resonance modes to extract diffusion parameters in the superconductor. Our work provides a general approach to study the quasi-particle dynamics in the superconducting resonator devices and design the photon-counting MKIDs.

## SUPPLEMENTARY MATERIAL

See supplementary material for derivation of equivalent resonance circuit, detailed description on the variable inductance model and Gaussian diffusion model.

## AUTHOR'S CONTRIBUTIONS

J. Hu and Q. He contributed equally to this work.

## ACKNOWLEDGMENTS

The devices were fabricated and tested at Southwest Jiaotong University. This work was supported in part by the National Natural Science Foundation of China (Grant Nos. 61871333, 11974290) and SRTP 202010613057.

## DATA AVAILABILITY

The data that support the findings of this study are available from the corresponding author upon reasonable request.

[1] J. Zmuidzinas, Annu. Rev. Condens. Matter Phys. **3**, 169 (2012).
[2] P. K. Day, H. G. LeDuc, B. A. Mazin, A. Vayonakis, and J. Zmuidzinas, Nature. **425**, 817 (2003).
[3] S. Doyle, P. Mauskopf, J. Naylon, A. Porch, and C. Duncombe, J. Low Temp. Phys. **151**, 530 (2008).
[4] J. Baselmans, J. Low Temp. Phys. **167**, 292 (2012).
[5] J. C. van Eyken, M. J. Strader, A. B. Walter, S. R. Meeker, P. Szypryt, C. Stoughton, K. OBrien, D. Marsden, N. K. Rice, Y. Lin, and B. A. Mazin, Astrophys. J. Suppl. Ser. **219**, 14 (2015).
[6] S. R. Golwala, C. Bockstiegel, S. Brugger, N. Czakon, P. Day, T. P. Downes, R. Duan, J. Gao, A. K. Gill, J. Glenn, S. R. Golwala, M. I. Hollister, A. Lam, H. G. LeDuc, P. R. Maloney, B. A. Mazin, S. G. McHugh, D. A. Miller, A. K. Mroczkowski, O. Noroozian, H. T. Nguyen, J. A. Schlaerth, S. R. Siegel, A. Vayonakis, P. R. Wilson, and J. Zmuidzinas, SPIE Proc. **8452**, 845205 (2012).
[7] A. Monfardini, L. Swenson, A.Bideaud, A. Cruciani, P. Camus, C. Hoffmann, F. Desert, S.Doyle, P. Ade, P. Mauskopf, C. Tucker, M. Roesch, S. Leclercq, K. F. Schuster, A. Endo, A. Baryshev, J. J. A. Baselmans, L. Ferrari6, S. J. C. Yates, O. Bourrion, J. Macias-Perez, C. Vescovi, M. Calvo, and C. Giordano, Astrophys. J. Suppl. Ser. **194**, 24 (2011).
[8] J. Hubmayr, J. Beall, D. Becker, H.-M. Cho, M. Devlin, B. Dober, C. Groppi, G. C. Hilton, K. D. Irwin, D. Li, P. Mauskopf, D. P. Pappas, J. V. Lanen, M. R. Vissers, Y. Wang, L. F. Wei, and J. Gao1, Appl. Phys. Lett. **106**, 073505 (2015).
[9] S. Rowe, E. Pascale, S. Doyle, C. Dunscombe, P. Hargrave, A. Papageorgio, K. Wood, P. Ade, P. Barry, A. Bideaud, T. Brien, C. Dodd1, W. Grainger, J. House1, P. Mauskopf, P. Moseley, L. Spencer, R. Sudiwala, C. Tucker1, and I. Walker, Rev. Sci. Instrum. **87**, 033105 (2016).
[10] J. Gao, M. Visser, M. Sandberg, F. Silva, S. Nam, D. Pappas, D. Wisbey, E. Langman, S. Meeker, B. Mazin, H. Leduc, J. Zmuidzinas, and K. Irwin, Appl. Phys. Lett. **101**, 142602 (2012).
[11] W. Guo, X. Liu, Y. Wang, Q. Wei, L. F. Wei, J. Hubmayr, J. Fowler, J. Ullom, L. Vale, M. R. Vissers, and J. Gao, Appl. Phys. Lett. **110**, 212601 (2017).
[12] R. Mezzena, M. Faverzani, E. Ferri, A. Giachero, B. Margesin, A. Nucciotti, A. Puiu, and A. Vinante, J. Low Temp. Phys. **199**, 73 (2020).
[13] N. Zobrist, G. Goiffard, B. Bumble, N. Swimmer, S. Steiger, M. Daal, G. Collura, A. Walter, C. Bockstiegel, N. Fruitwala, I. Lipartito, and B. A. Mazin, Appl. Phys. Lett. **115**, 213503 (2019).
[14] J. D. Parrianen, A. Papageorgiou, S. Doyle, and E. Pascale, J. Low Temp. Phys. **193(3-4)**, 113 (2018).
[15] P. Visser, S. A. H. de Rooij, V. Murugesan, D. J. Thoen, and J. J. A. Baselmans, arXiv:2103.06723 (2021).
[16] O. Quaranta, T. Cecil, L. Gades, B. Mazin, and A. Miceli, Supercond. Sci. Technol. **26**, 105021 (2013).
[17] B. Mazin, B. Bumble, P. Day, M. Eckart, S. Golwala, J. Zmuidzinas, and F. Harrison, Appl. Phys. Lett. **89**, 222507 (2006).
[18] G. A.Vardulakis, S.Withington, and D. J.Goldie, SPIE Proc. **5499**, 348 (2004).

[19] R. Barends, J. J. Baselmans, S. J. Yates, J. R. Gao, J. N. Hovenier, and T. M. Klapwijk, Phys. Rev. Lett. **100(25)**, 257002 (2008).
[20] C. Wang, Y. Y. Gao, I. M. Pop, U. Vool, C. Axline, T. Brecht, R. W. Heeres, L. Frunzio, M. Devoret, G. Catelani, L. Glazman, and R. J. Schoelkopf, Nature communications. **5(1)**, 1 (2014).
[21] L. Grünhaupt, N. Maleeva, S. T. Skacel, M. Calvo, F. Levy-Bertrand, A. Ustinov, H. Rotzinger, A. Monfardini, G. Catelani, and I. M. Pop, Phys. Rev. Lett. **121(11)**, 117001 (2018).
[22] R. P. Budoyo, J. B. Hertzberg, C. Ballard, K. D. Voigt, Z. Kim, J. R. Anderson, C. J. Lobb, and F. C. Wellstood, Phys. Rev. B. **93(2)**, 024514 (2016).
[23] C. Bellenghi, L. Cardani, N. Casali, I. Colantoni, A. Cruciani, G. Pettinari, and M. Vignati, J. Low Temp. Phys. **199**, 639 (2020).
[24] A. Kozorezov, A. Volkov, J. Wigmore, A. Peacock, A. Poelaert, and R. Hartog, Physical Review B. **61(17)**, 11807 (2000).
[25] T. Guruswamy, D. Goldie, and S. Withington, Supercond. Sci. Technol. **27**, 055012 (2014).
[26] T. Guruswamy, D. Goldie, and S. Withington, Supercond. Sci. Technol. **28**, 054002 (2015).
[27] S. Hsieh and J. Levine, Phys. Rev. Lett. **20**, 1502 (1968).
[28] M. Pyle, P. Brink, B. Cabrera, J. Castle, P. Colling, C. Chang, J. Cooley, T. Lipus, R. Ogburn, and B. Young, Nucl. Instrum. Methods. **A559**, 405–407 (2006).
[29] S. Friedrich, K. Segall, M. Gaidis, C. Wilson, D. Prober, A. Szymkowiak, and S. Moseley, Appl. Phys. Lett. **71**, 3901 (1997).
[30] R. Barends, S. Vliet, J. Baselmans, S. Yates, J. Gao, and T. Klapwijk, Physical Review B. **79**, 020509 (2009).
[31] N. Vercruyssen, R. Barends, T. Klapwijk, J. Muhonen, M. Meschke, and J. Pekola, Appl. Phys. Lett. **88**, 062509 (2011).
[32] P. Visser, J. Baselmans, S. Yates, P. Diener, A. Endo, and T. Klapwijk, Appl. Phys. Lett. **100**, 162601 (2012).
[33] A. Fyhrie, P. Day, J. Glenn, H. Leduc, C. McKenney, J. Perido, and J. Zmuidzinas, J. Low Temp. Phys. **199**, 688 (2020).
[34] J. Gao, J. Zmuidzinas, A. Vayonakis, P. Day, B. Mazin, and H. Leduc, J. Low Temp. Phys. **151**, 557 (2008).
[35] A. Porch, P. Mauskopf, S. Doyle, and C. Dunscombe, IEEE Trans. Appl. Supercond. **15**, 552 (2005).
[36] Y. Wang, P. Zhou, L. Wei, H. Li, B. Zhang, Z. Miao, W. Qiang, Y. Fang, and C. Cao, J. Appl. Phys. **114**, 153109 (2013).
[37] J. Gao, Ph.D. thesis, Caltech. (2008).
[38] B. Mazin, Ph.D. thesis, Caltech. (2004).